\documentclass{jpsj2}

\def\runauthor{Online-Journal Subcommittee}

\title{%
Observation of lattice distortion in the low-dimensional quantum spin system TiOBr by synchrotron x-ray diffraction   - Evidence of spin-Peierls transition? --}

\author{%
Tomoyuki Sasaki, Masaichiro Mizumaki$^1$, Kenichi Kato$^{1,2}$, Yasuo Watabe, Yoshiki Nishihata, Masaki Takata$^{1,2}$ and Jun Akimitsu
}

\inst{%
Department of Physics, Aoyama-Gakuin University, Sagamihara, Kanagawa 229-8558\\
$^1$Japan Synchrotron Radiation Reserch Institute (JASRI), SPring-8,Mikazuki-cho, Sayo-gun, Hyogo 679-519\\
$^2$CREST, Japan Science and Technology Corporation (JST), Kawaguchi, Saitama 332-0012
}

\recdate{December 23, 2004 }

\abst{%
We report here on physical properties and lattice distortion in the new quasi-one-dimensional spin system TiOBr by x-ray diffraction experiment using synchrotoron radiation. Comparing the crystal structure between TiOBr and TiOCl, the difference of $b$-axis length plays a key role for physical properties in both systems. By using single crystal of TiOBr, superlattice reflections were observed at (0 1.5 0) and (0 2.5 0). The temperature dependence of superlattice reflections imply the first order transition at $T_{c1}$=27K. We analyzed the data using the simple dimerization model.
}

\kword{%
Spin-Peierls transition, lattice distortion, x-ray synchrotoron radiation, crystal structure, dimerization
}

\begin{document}
\maketitle


@$S$ = 1/2 Quantum Heisenberg Antiferromagnetic Chain (QHAC) is one of the best candidate to investigate a role of quantum spin fluctuations. A small perturbation to QHAC causes a drastic change of the ground state with spontaneous broken symmetry. The spin-Peierls state being coupled with phonons results in a new disordered ground state with a finite energy gap for the excitation spectrum, accompanying a spontaneous dimerization of the spin and the lattice.

 The inorganic compound CuGeO$_3$ has proven to be a first inorganic spin-Peierls system and a lot of works, both theoretical and experimental, has been devoted to this material \cite{1, 2}. After the discovery of this material, impurity effect on the spin-Peierls compound was examined \cite{3}. Neutron scattering experiments on CuGe$_{0.993}$Si$_{0.007}$O$_3$ demonstrated the coexistence of the Bragg spot intensities for both lattice dimerization and antiferromagnetic (AF) orderings at low temperature \cite{4}. The two Bragg spots are resolution limited, which implies that the two orderings are truly long ranged. The result was very surprising since the two orderings had been shown to be exclusive in homogeneous systems. A theoretical proposal was presented for understanding of this remarkable phenomenon using phase Hamiltonian technique at $T$ = 0K \cite{5}. This fact suggests that a new phenomenon can be created by a new perturbation such as orbital and charge degrees of freedom.
 
 Recently, the layered compound TiOCl has been suggested to be a new spin-Peierls system with one-dimensional (1D) orbital ordered array, by which 1D chain composes \cite{6}. This material has long been believed to be a $S$=1/2 two-dimensional antiferromagnet because of the layered structure and having a very small and temperature independent magnetic moment \cite{7}, which suggests that the ground state is the Resonating-Valence-Bond (RVB) state in the Mott-insulator.
 
 The structure of TiOCl is of FeOCl type, where the TiO$_4$Cl$_2$ bilayers separate from each other along c-axis. The distorted TiO$_4$Cl$_2$ octahedra build an edge-shared network. The nearest-neighbor Ti ions form a staggered site by $a/2$ and $b/2$ between each layer along the $a$- and $b$ - axis. The next-nearest-neighbor Ti ions form an edge-shared Ti ion along $b$ - axis in the same plane. The most important exchange path is direct $t_{\rm 2g}$ orbital overlap, rather than superexchange interaction. In this case, the chains are expected to form along $a$- or $b$-axes [see Fig.1 b) and c) in Ref.\cite{6}]. However, by comparing ESR measurements to angular overlap model (AOM) \cite{8} and by generalized gradient approximation (GGA) which suggests that TiOCl is subject to large orbital fluctuations driven by the electron-phonon coupling \cite{9}, a uniform $S$ = 1/2 chains are suggested to form along $b$-axis because of the occupied Ti dxy orbital. The temperature dependence of susceptibility can be fitted using $S$ = 1/2 Bonner-Fisher curve with an exchange constant of $J$ = 660K, and can be observed a sharp drop at $T_{\rm c1}$ = 67K and a pronounced kink at $T_{\rm c2}$ = 95K. It was also observed by the NMR measurement that the opening of the pseudo-spin gap at $T^*$ $\sim$ 135K \cite{10}. The pseudo-spin gap energy was evaluated to be $\Delta_{\rm NMR} \sim$ 430K by NMR \cite{10} and 2$\Delta_{opt} \sim$ 430K by optical measurement below $T_{\rm c1}$ \cite{11, 12}. In spite of such a large pseudo-spin gap, the phase transition is suppressed by a strong coupling between spin and lattice degrees of freedom and orbital fluctuation, and as the result, is realized the spin gap state at $T_{\rm c1}$. TiOCl is possible to be an 1D system by the formation of the orbital ordering along $b$-axis, and finally the system causes the spin-Peierls instability driven by the electron-phonon interaction.
 
\begin{figure}[htbp]
\begin{center}
\includegraphics[width=8.5cm]{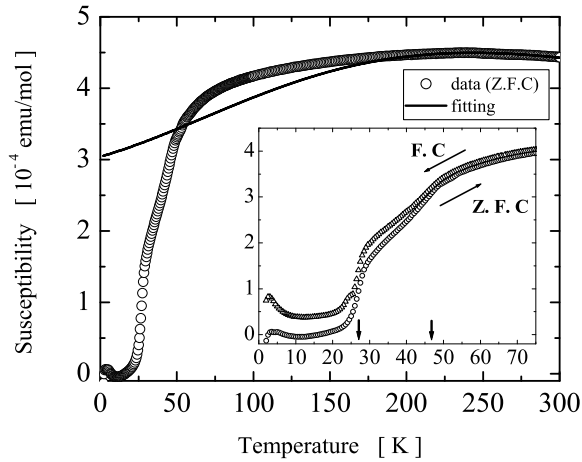}
\end{center}
\caption{Temperature dependence of susceptibilities for polycrystalline TiOBr in a magnetic field of 5.5 T without Curie term subtraction. Arrows indicate magnetic transition temperatures $T_{\rm c1}$ and $T_{\rm c2}$. The solid line indicates the fitting result with Bonner-Fisher curve. The inset is emphasized at low temperature region.}
\label{f1}
\end{figure}

 It is expected that the 1D nature in TiOBr is weaker than that in TiOCl, because the lattice parameter ratio ($a$-axis/$b$-axis) in TiOBr is smaller than that in TiOCl. Actually, physical properties of 1D nature obtained by the temperature dependence of susceptibility in TiOBr are proportionally weaker than that in TiOCl \cite{13}. Although the measurement for possible structural phase transition has been examined by x-ray and neutron powder diffraction, it has not yet been found. 
 
 In this letter, we present the observation of the superlattice reflections probably due to the spin-Peierls transition by the SR x-ray diffraction in the low dimensional quantum spin system TiOBr. 
 
 Polycrystalline and single crystal samples of TiOBr were synthesized from Ti, TiO$_2$ and TiBr$_4$ by the chemical vapor transport method \cite{6, 13, 14}. The precursors of TiOBr were prepared from a pellet of Ti/TiO$_2$ (1:2 by mol) and sextuple TiBr$_4$. In the evacuated sealed quartz tube, there were a pellet of TiBr$_4$ at the cooled zone and of Ti/ TiO$_2$ at the hot zone, where both were heated up to 700Ž with a gradient at 100Ž within 20 cm. It takes 3 days to cool down from 700Ž to 450Ž. The excess TiBr$_4$ was carried away at cooled zone. The rectangle single crystal ($2 \times 10 \times 0.04$ mm) was grown in $ab$ - plane, and the major axis corresponds to $b$-axis.
 
 The polycrystalline samples were characterized by the x-ray powder diffraction, which showed a small amount of Ti$_2$O$_3$ phase. To characterize the single crystal, the x-ray powder diffraction, EPMA and Laue technique were employed, which shows no evidence of impurity phase and its composition being stoichiometric.
 
\begin{figure}[htbp]
\begin{center}
\includegraphics[width=8cm]{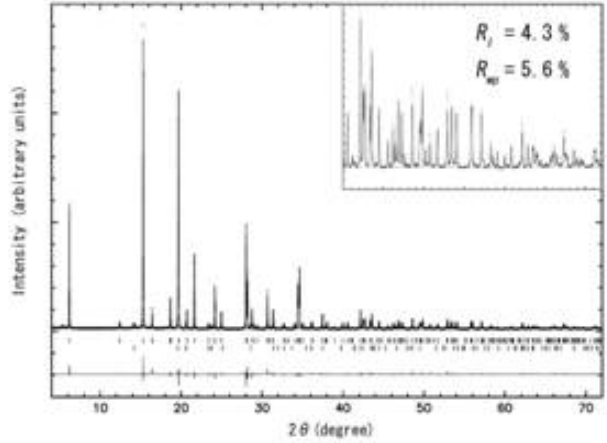}
\end{center}
\caption{Rietveld analysis of the x-ray diffraction pattern for polycrystalline TiOBr being fitted for the orthorhombic $Pmmn$ model. The data were taken at 300K, in which several weak impurity peaks were found and identified as Ti$_2$O$_3$.}
\label{f2}
\end{figure}
\begin{table}[htbp]
\vspace{2mm}
\label{t2}
(a)
\begin{center}
\begin{tabular}{ccccccccc} 
\hline
\hline

 &  & $x$ & & $y$ && $z$ && $B$(\AA) \\

\hline
\hline 

 Ti &  & 0.25(0) && 0.75(0) && 0.10951(7) && 0.33(1) \\

 O &  & 0.25(0) && 0.25(0) && 0.95151(27) && 0.22(5) \\

 Br &  & 0.25(0) && 0.25(0) && 0.32723(6) && 1.18(2) \\

\hline
\hline
\end{tabular}
\end{center}
{\tiny
Space Group $Pmmn$(59:2) $Z=2$\\
}
\\
(b)
\begin{center}
\begin{tabular}{ccccc}
\hline
\hline

 &  & $a$ & $b$ & $c$  \\

\hline
\hline

300K &  & 3.78458(2) & 3.48528(2) & 8.52520(5) \\
37K &  & 3.78014(51) & 3.46647(7) & 8.49861(69) \\
12K &  & 3.78307(36) & 3.46556(9) & 8.49643(57) \\

\hline
\hline
\end{tabular}
\end{center}
{\tiny
The factor of parameter are \AA
}
\caption{(a) Atomic parameters for TiOBr obtained from synchrotron x-ray diffraction at 300K. The space group is $Pmmn$. (b) Lattice parameters at 300, 37 and 12K.}
\end{table}
 
 Magnetic susceptibility was measured using a DC SQUID magnetometer (MPMSR2, Quantum Design) at 5.5 T. For precise crystal structure determination, SR x-ray powder diffraction experiments were carried out by large Debye-Scherrer camera installed at BL02B2, SPring-8 \cite{15}. The exposure time of x-ray was 160 min. The x-ray patterns were measured at 300K, 37K and 12K. The wavelength of incident x-ray was 0.922\AA. The obtained powder data were analyzed by the Rietveld method. To determine the lattice distortion, SR single crystal diffraction experiments were carried out on the four-circle diffractometer at BL46XU, SPring-8. The x-ray wavelength using Si (111) monochromators was 1.0332\AA. The size of single crystal is about $1.5 \times 6 \times 0.02$mm$^3$, which mountains onto refrigerator with the (h k 0) reciprocal plane parallel to the $\chi$-axis.

In Fig. 1, the susceptibility data for TiOBr were obtained from the polycrystalline samples. Although the temperature dependence of susceptibility in TiOBr resemble that in TiOCl \cite{6, 8}, we observed a small hysteresis curve in TiOBr due to a different cooling process (field cooling and zero field cooling processes) below 280K. The data in Fig. 1 are subtracted the Van-Vleck susceptibility ($\chi_0$) and the Curie term of Ti$^{3+}$ below 20K from the raw data. The Curie term is originated from the paramagnetic (PM) component in Ti$_2$O$_3$ due to the deterioration and the unpaired free $S$ = 1/2 spins with the finite chain effects. The Curie term is order of 0.8$\%$/Ti$^{3+}$, and the $\chi_0$ is $1.3\times 10^{-5}$ emu/mol. The data above 220K are fitted to the Bonner-Fisher curve \cite{16} with the nearest -neighbor exchange $J$ = 188K and the g factor of 1.752. The $g$ factor was obtained from the ESR measurement \cite{13}. The data between 10 and 20K are fitted to the function of Troyer ${et}$ ${al}$. who suggested that the susceptibility of the spin gap system decreases as $a T^{-1/2}$exp(-$\Delta/k_{\rm B}T$) at sufficiently low temperature ($T << \Delta/k_{\rm B}$) \cite{17}, which gives the energy gap $\Delta/k_{\rm B}$ = 149K. The data shown in the inset of Fig. 1 features a rapid drop of the susceptibility at $T_{\rm c1}$ = 27K and $T_{\rm c2}$ = 46K, which are defined by the value of the differentiated susceptibility. In addition, we observed the third anomaly around 10K ($T_{\rm c3}$) at which susceptibility increases again with decreasing temperature. The ratio $2\Delta/k_{\rm B}T_{\rm c1,c2}$ = 6.5$\sim$11 is smaller than $2\Delta/k_{\rm B}T_{\rm c1,c2}$ = 9$\sim$13 of TiOCl \cite{10}, and these values are much larger than 3.5 expected from BCS theory.
\begin{table*}[tb]
\begin{center}
\begin{tabular}{ccccccccccccccc} 
\hline
\hline

 &  & $a$ [\AA] & & $b$ [\AA] && $c$ [\AA] && $T_{\rm c1}$ && $T_{\rm c2}$ && $J$ [K] && $\Delta/k_{\rm B}$ \\

\hline
\hline 

 TiOBr &  & 3.785 && 3.485 && 8.525 && 27 && 47 && 188 && 149\\

 TiOCl &  & 3.789 $^{8)}$ && 3.365 $^{8)}$ && 8.060 $^{8)}$ && 67 $^{6)}$ && 95 $^{6)}$ &&660 $^{6)}$ && 440 $^{10)}$ \\

\hline
\hline
\end{tabular}
\begin{tabular}{cccccccccccccccccc} 
\hline
\hline

 &&&  & Ti-Ti($a$) [\AA] &&& Ti-Ti-Ti($a$) [$^\circ$] &&& Ti-Ti($b$) [\AA] &&& Ti-Ti-Ti ($b$) [$^\circ$] & \\

\hline
\hline 

  &&&  & 3.1787 &&& 73.07 &&& 3.4853 &&& 180 & \\

  &&&  & 3.1811  &&& 73.10 &&& 3.3650 &&& 180 & \\

\hline
\hline
\end{tabular}
\end{center}
\caption{The summarized physical properties in TiOBr and TiOCl. The lower column shows the distances and angles between the nearest-neighbor Ti atoms in zig-zag chain along $a$-axis and linear chain along $b$-axis suggested by Seidel ${et}$ ${al.}$ \cite{6}.}
\end{table*}


Figure 2 shows the Rietveld analysis of the SR x-ray diffraction pattern for polycrystalline TiOBr  at 300K for which the orthorhombic $Pmmn$ model was fitted. Table.1 shows atomic parameters at 300K and its lattice parameters at 300, 27 and 12K. It was finally confirmed that the TiOBr has a same crystal structure with that of TiOCl \cite{18}. Our x-ray diffraction analysis shows that the structural difference between TiOBr and TiOCl is mainly its lattice parameter. From the temperature dependence of the structural analysis, superlattice reflction due to lattice distortion was not observed between the lowest temperature and room temperature. This is probably due to the weak intensity of the superlattice reflection in the powder reflection data. However, lattice parameters decrease with decreasing temperature, in particular $b$-axis drastically changes, which confirms the development of intra-interaction with decreasing temperature.	

\begin{figure}[tbp]
\begin{center}
\includegraphics[width=7cm]{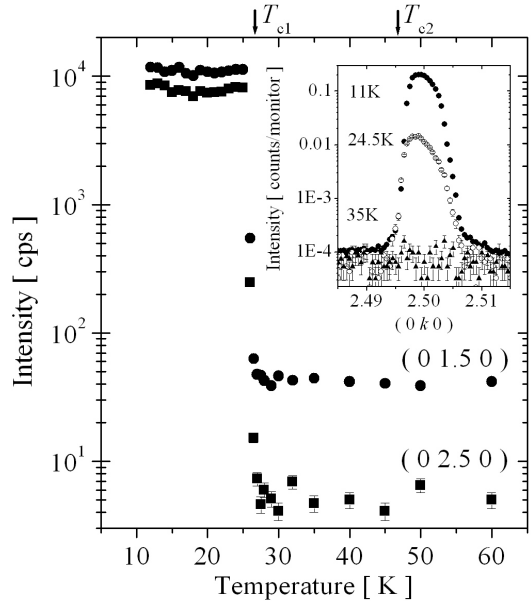}
\end{center}
\caption{Temperature dependence of the superlattice reflections at (0 1.5 0) and (0 2.5 0) on heating process. The inset shows the peak profiles of the (0 2.5 0) reflection.}
\label{f3}
\end{figure}

Table 2 shows the summary of the physical properties both in TiOBr and TiOCl. As shown in Table 2, roughly speaking, the value of the physical properties in TiOBr are 30`50$\%$ smaller than these of TiOCl. This corresponds that $b$-axis in TiOBr is longer than that in TiOCl, namely the intra-chain exchange in TiOBr is weaker than that in TiOCl. Moreover, although there is no significant difference of Ti-Ti distance along $a$-axis between TiOCl and TiOBr, the Ti-Ti distance along $b$-axis is significantly different between them.

Seidel ${et}$ ${al.}$ interpreted the abrupt drop of the susceptibility at $T_{\rm c1}$(=67K) as a transition into a spin-Peierls state \cite{6} in TiOCl. We have clearly observed the broad anomalies in heat capacity at $T_{\rm c1}$ and $T_{\rm c2}$, and change of the relaxation rate in $\mu$SR and of ESR signal under high magnetic field at $T_{\rm c1}$, $T_{\rm c2}$ and $T_{\rm c3}$\cite{19, 20}.

What is happened at $T_{\rm c1}$ in this system? In order to search the superlattice reflections, we have performed a precise x-ray scattering measurement by using the single crystal of TiOBr. Consequently, we observed superlattice reflections at (0 1.5 0) and (0 2.5 0) by SR x-ray diffraction, which evidenced the lattice distortion in single crystal of TiOBr. The temperature dependence of intensities of superlattice reflections (0 1.5 0) and (0 2.5 0) are shown in Fig.3. With increasing temperature, the intensities gradually decreased, and suddenly dropped at $T_{\rm c1}$. Although it is not shown in Fig.3, this phase transition has thermal hysteresis around $T_{\rm c1}$, suggesting that this transition is of first order. Although the intensities above $T_{\rm c1}$ are small, the intensities seem to still survive between $T_{\rm c1}$ and $T_{\rm c2}$. 

\begin{figure}[htbp]
\begin{center}
\includegraphics[width=8.5cm]{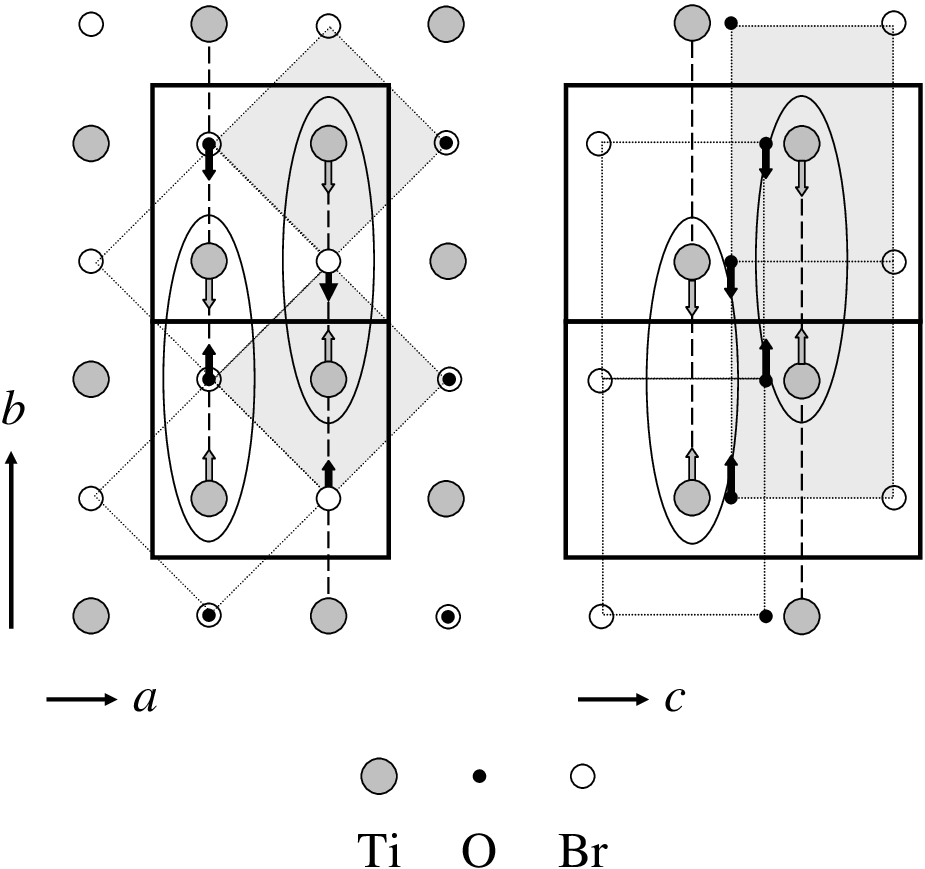}
\end{center}
\caption{Atomic shifts at 11K in TiOBr. The figure shows the model in which Ti and O atoms are shifted. Bold frame indicates the unit cell at high temperature, and the new unit cell are doubled along $b$-direction. Dashed frame indicates TiO$_4$Br$_2$ octahedra, of which color shows the different layer each other. Arrows indicate the directions of shifts.}
\label{f4}
\end{figure}

Degree of lattice distortion ($\delta$) was estimated from the superlattice reflections. The intensities of (0 1.5 0) and (0 2.5 0) reflections were normalized by (0 2 0) main reflection. Since only two reflections were observed, we employed the simpest model in which only Ti atoms are shifted. This model, unfortunately, can not explain the intensity ratio between (0 1.5 0) and (0 2.5 0). Next model is that all atoms along $b$-axis are shifted with same value. In order to simplify the model we employ the three types of shifts, i.e. (a) three kind of atoms, (b) two kind of atoms, and (c) one kind of atom, and totally six type models are analyzed for this estimation. Fig.4 shows that Ti and O atoms shifts along $b$-axis for example. This simple estimation supports two models; one is that Ti and O atoms are shifted with $\delta_{\rm Ti,O} = 0.0022 \pm 0.0002$, and the other is that only O atom is shifted ($\delta_{\rm O}=0.0161 \pm 0.0016$). The unit is given as doubled unit cell ($b'=2b$) in reciprocal lattice unit. In both cases, the shift of O atoms plays a key role for the estimation. Assuming that 1D chains are aligned along $b$-axis, $d_{\rm xy}$ orbital in Ti atom should be occupied by $3d$ electron \cite{6}, i.e. Ti atoms centered in TiO$_2$Br$_2$ ribbon are arrayed along $b$-axis (see Fig.4 right picture). 

In summary, we have reported the magnetic properties and structural analysis in TiOBr. From the structural analysis, we found that TiOBr has a similar crystal structure with that of TiOCl except a large difference of $b$-axis length, as shown in Table 2. However, we emphasize here that the $b$-axis difference substancially affect the one dimensionality for both systems. From the single crystal x-ray diffraction measurements, we clearly observed the superlattice reflections along $b$-axis, which is of first order transition. We tried to fit the intensities of the superlattice reflection to the simple dimerization model as shown in Fig.4. More detailed experiment for the crystal structure determination is now in progress which will be published in near future.

We would like to thank S. Kimura (Spring-8) for technical supports, and T. Yokoo (KEK) for his interest in this work and for fruitful discussion. Aoyama-Gakuin group was partly supported by 21st COE program, "High-Tech Reserch Center" Project for Private Universities: matching fund subsidy from MEXT (Ministry of Education, Sports, Culture, Science and Technology), 2002-2004 and Grant-in-Aid for Scientific Research on Priority Area Ministry of Education, Sports, Culture, Science and Technology of Japan. The synchrotron radiation experiments were performed at the BL02B2 and BL46XU (R04B46XU-0020N) in the SPring-8 with the approval of the Japan Synchrotron Radiation Research Institute (JASRI)



\begin{thebibliography}{99}

\bibitem{1} M. Hase, I. Terasaki and K. Uchinokura: Phys. Rev. Lett. {\bf 70} (1993) 3651

\bibitem{2} M. Nishi, O. Fujita and J. Akimitsu: Phys. Rev. B {\bf 50} (1994) 6508

\bibitem{3} M. Hase, I. Terasaki, Y. Sasago, K. Uchinokura and H. Obara: Phys. Rev. Lett. {\bf 71} (1993) 4059

\bibitem{4} L. P. Regnault, J. P. Renard, G. Dhalenne and A. Revcolevschi: Europhys. Lett. {\bf 32} (1995) 579

\bibitem{5} H. Fukuyama, T. Tanimoto and M. Saito: J. Phys. Soc. Jpn. {\bf 65} (1996) 1182

\bibitem{6} A. Seidel, C. A. Marianetti, F. C. Chou, G. Ceder and P. A. Lee: Phys. Rev. B {\bf 67} (2003) 020405(R)

\bibitem{7} R. J. Beynon and J. A. Wilson: J. Phys.:Condens. Matter {\bf 5} (1993) 1983

\bibitem{8} V. Kataev, J. Baier, A. Moller, L. Jongen, G. Meyer and A. Freimuth: Phys. Rev. B {\bf 68} (2003) 140405(R)

\bibitem{9} T. Saha-Dasgupta, R. Valenti, H. Rosner and C. Gros: Europhys. Lett. {\bf 67} (2004) 63

\bibitem{10} T. Imai and F. C. Chou: cond-mat/0301425 (2003)

\bibitem{11} P. Lemmens, K. Y. Choi, G. Caimi, L. Degiorgi, N. N. Kovaleva, A. Seidel and F. C. Chou: Phys. Rev. B {\bf 70} (2004) 134429

\bibitem{12} G. Caimi, L. Degiorgi, N. N. Kovaleva, P. Lemmens and F. C. Chou: Phys. Rev. B {\bf 69} (2004) 125108

\bibitem{13} C. Kato, K. Tsuchida, Y. Kobayashi and M. Sato: J. Phys. Soc. Jpn. {\bf 74} (2005) 473

\bibitem{14} H. G. v. Schnering, M. Collin and M. Hassheider: Z. Anorg. Allg. Chem. {\bf 383} (1972) 137

\bibitem{15} E. Nishibori, M. Takata, K. Kato, M. Sakata, Y. Kubota, S. Aoyagi, Y. Kuroiwa, M. Yamakata and N. Ikeda: Nuclear Instruments and Methods in Physics Research A, 467 (2001) 1045

\bibitem{16} W. E. Hatfield: J. Appl. Phys. {\bf 52} (1981) 1985

\bibitem{17} M. Troyer, H. Tsunetsugu and D. Wurtz: Phys. Rev. B {\bf 50} (1994) 13515

\bibitem{18} H. Schafer, F. Wartenpfuhl and E. Weise: Z. Anorg. Allg. Chem. {\bf 295} (1958) 268

\bibitem{19} T. Sasaki, Y. Nishihata, H. Takagiwa, S. Kuroiwa, K. Ohishi, W. Higemoto, A. Kouda, Y. Kadono and J. Akimitsu: to be published

\bibitem{20} S. Kimura, T. Sasaki, K. Kindo and J. Akimitsu: to be published

\end{thebibliography}
\end{document}